# Homopolar Chemical Bonds Induce In-Plane Anisotropy in Layered Semiconductors

*Jieling Tan[1], Jiang-Jing Wang[1]\*, Hang-Ming Zhang[1], Han-Yi Zhang[1], Heming Li[1,2], Yu Wang[2], Yuxing Zhou[3], Volker L. Deringer[3]\*, Wei Zhang[1]\**

[1]Center for Alloy Innovation and Design (CAID), State Key Laboratory for Mechanical Behavior of Materials, Xi'an Jiaotong University, Xi'an, 710049, China

[2]School of Physics, Xi'an Jiaotong University, Xi'an, 710049, China

[3]Inorganic Chemistry Laboratory, Department of Chemistry, University of Oxford, Oxford, OX1 3QR, UK

\*Emails: j.wang@mail.xjtu.edu.cn, volker.deringer@chem.ox.ac.uk, wzhang0@mail.xjtu.edu.cn

## Abstract:

Main-group layered binary semiconductors, in particular, the III–VI alloys in the binary Ga–Te system are attracting increasing interest for a range of practical applications. The III–VI semiconductor, monoclinic gallium monotelluride (m-GaTe), has been recently used in high-sensitivity photodetectors/phototransistors and electronic memory applications due to its anisotropic properties yielding superior optical and electrical performance. Despite these applications, the origin of such anisotropy, namely the complex structural and bonding environments in GaTe nanostructures remain to be fully understood. In the present work, we report a comprehensive atomic-scale characterization of m-GaTe by state-of-the-art element-resolved atomic-scale microscopy experiments, enabling a direct measure of the in-plane anisotropy at the sub-Angstrom level. We show that these experimental images compare well with the results of first-principles modeling. Quantum-chemical bonding analyses provide a detailed picture of the atomic neighbor interactions within the layers, revealing that vertical Ga–Ga homopolar bonds get stronger when they are distorted and rotated, inducing the strong in-plane anisotropy. Beyond GaTe, using a systematic screening over the Materials Project database, we identify four additional low-symmetric layered crystals with similar distorted tetrahedral patterns, namely, GeP, GeAs, SiP, and SiAs. We thereby confirm the homopolar-bond-induced anisotropy is a more generic feature in these layered van-der-Waals materials.

# 1. Introduction

Main-group chalcogenides, including the III–VI semiconductors in the binary Ga–Te system, are of increasing interest with regard to structural and chemical properties as well as a range of practical applications.[1-4] The monotelluride, GaTe, occurs in two different polymorphs: a complex monoclinic structure ("m-GaTe" in the following)[5] and a hexagonal one ("h-GaTe").[6] The two polymorphs both contain layered-like building blocks with a stacking sequence of Te–Ga–Ga–Te, connected in the third dimension via van der Waals (vdW) interactions. The m-phase is energetically more favorable in the bulk, whereas the h-phase has a lower energy in few-layer ultrathin film form.[7] Approaching the two-dimensional limit, both phases can be obtained by proper phase engineering under different experimental conditions.[8-10] Similar to other h-phases of III–VI semiconductors, in particular h-InSe and h-GaSe which have been shown to exhibit superplastic deformability,[11-15] the h-phase of GaTe has also been predicted to be a potential candidate for next-generation deformable or flexible electronics.[12-13] As regards the m-phase of GaTe, its structural complexity leads to excellent optical[16-18] and electrical[19-20] properties, enabling high-sensitivity photodetectors/phototransistors[21-25] and anisotropic non-volatile electronic memory applications.[19]

Besides m-GaTe, other emerging two-dimensional materials with in-plane anisotropy include phosphorene,[26-27] transition-metal chalcogenides,[28] group III– or IV chalcogenides,[29-33] and group IV pnictogenides.[34-36] In fact, the anisotropy-based applications intrinsically originated from the large complexity and distortion of the two-dimensional layered materials, specifically, the low in-plane symmetry of their crystal structures. Such structural distortion results in different responses to external stimuli (e.g., optical, electrical, and mechanical properties) along different crystallographic directions. For most anisotropic materials, e.g., SnSe,[37] 1T'-MoTe$_2$,[38] and GeS$_2$,[39-40] homopolar bonds are absent and the structural anisotropy mostly results from the distortion of heteropolar bonds. For instance, in monoclinic GeS$_2$, the major local motifs are edge- and corner-sharing tetrahedral [GeS$_4$] units, and its structural anisotropy stems from tilted heteropolar Ge–S bonds of the corner-sharing GeS$_4$ tetrahedra along the *x*-axis.[39]

In contrast, there is a considerable amount of homopolar (Ga–Ga) bonds in m-GaTe, and it remains unclear whether the structural anisotropy stems from homopolar or heteropolar bonds. In the present work, we provide atomic-scale structural characterization experiments to visualize and measure the anisotropy at the sub-Angstrom level via Cs-corrected scanning transmission electron microscopy (STEM) and energy-dispersive X-ray (EDX) mapping experiments. Complementing these atomic-scale imaging experiments, we quantify the chemical-bonding configurations that are present m-GaTe using density functional theory (DFT) computations and orbital-based bonding analysis. The combined experimental and theoretical efforts elucidate a more generic homopolar-bond-induced anisotropy in m-GaTe and related layered semiconductors.

# 2. Results and discussion

We carried out DFT calculations with projector augmented wave (PAW) pseudopotentials,[41-42] Perdew–Burke–Ernzerhof (PBE) functional,[43] and Grimme D3 method[44] for vdW correction using VASP.[45] The relaxed layered-like structures of m-GaTe and h-GaTe are shown in **Figure 1**. The computed lattice parameters are $a$ = 17.73 Å, $b$ = 10.54 Å, $c$ = 4.08 Å, $\alpha = \beta$ = 90°, $\gamma$ = 104.4° for m-GaTe, and $a = b$ = 4.12 Å, $c$ = 16.96 Å, $\alpha = \beta$ = 90°, $\gamma$ = 120° for h-GaTe, consistent with previous experimental results.[5, 46] The first question regarding the GaTe polymorphs is whether the structures should in fact be viewed as one-, two-, or three-dimensionally extended; this is in analogy to previous studies of the covalent and vdW interactions in crystalline $Sb_2Se_3$.[47] For m-GaTe, the aforementioned Ga–Ga homopolar contacts play a crucial role in addressing this question: hypothetically, if one disregards the in-plane (i-) Ga–Ga contacts, the structure appears to consist of one-dimensional "nanowires" extending along the [001] direction (dashed blue box in Figure 1a). Such 1D nanowires have been termed as pseudo-1D materials in ref. [16], in analogy with elemental Te.[48-49] However, the interatomic distance between the two Ga atoms in m-GaTe is 2.45 Å, much shorter than the heteropolar Ga–Te bond (2.69 Å). Hence, the i-Ga–Ga contact should be regarded as a chemical bond, which leads to an alternative description as a 2D material (green shading in Figure 1a). For the other polymorph, h-GaTe, the structure clearly appears as extending in 2D, but there are again two different ways of viewing it based on whether the homopolar Ga–Ga contacts are included or not (green and blue in Figure 1b, respectively). In the following, we will quantify the importance of homopolar i-Ga–Ga bonds in shaping the anisotropic properties of m-GaTe.

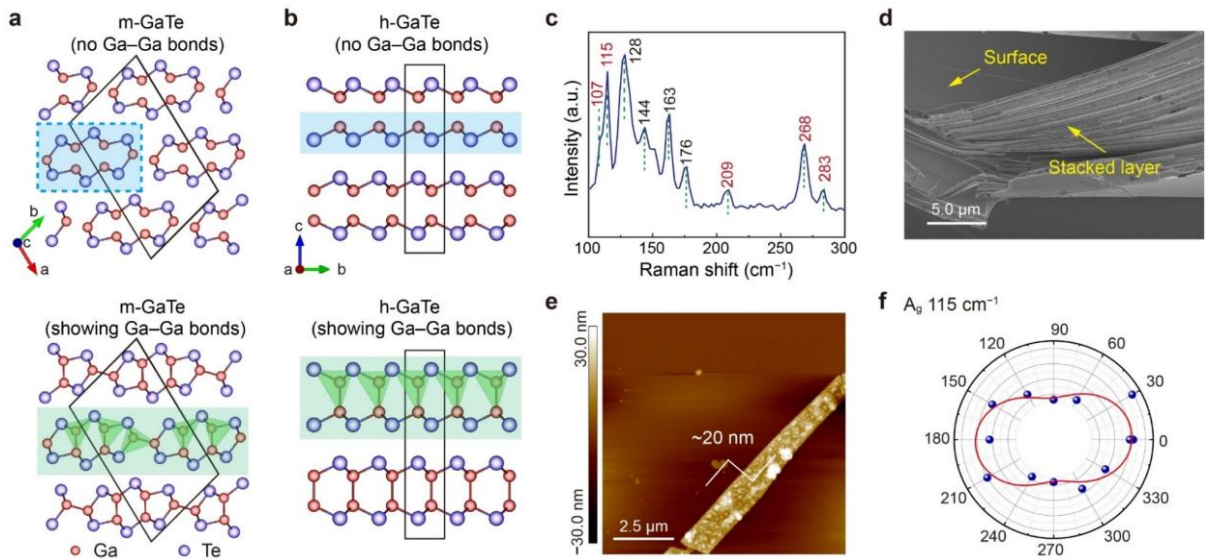

**Figure 1. Crystal structures and characterization of single-crystalline GaTe**. a–b) Crystal structures of the m- and h-phases without (blue) or with (green) showing Ga–Ga homopolar bonds. Red (purple) spheres indicate Ga (Te) atoms, respectively. Green tetrahedra highlight local Ga coordination environments in both structures with Ga–Ga bonds shown. The box with the blue dashed line marks the structural unit of m-GaTe. c) Raman spectrum of bulk m-GaTe, where the $A_g$ modes are labeled in red font. d) SEM image of a typical exfoliation of m-GaTe sheets. e) AFM image of an exfoliated m-GaTe flake which is ~20 nm thick. f) Raman intensity polar plot and fitting line (solid line) of the $A_g$ mode at 115 cm$^{-1}$ as a function of rotation angle for an m-GaTe thin flake. The laser excitation wavelength was 532 nm.

First, we obtained a single-crystal sample of m-GaTe via chemical vapor transport (see the X-ray diffraction pattern in Figure S1), and the measured Raman spectrum is consistent with the monoclinic structure of the sample (Figure 1c). In the Raman spectrum, the highest intensity peak at 128 cm$^{-1}$ indicates van der Waals interactions,[50] and the five $A_g$ modes[51] (at 107, 115, 209, 268, and 283 cm$^-$ [1]) are related to the in-plane vibrations. The scanning electron microscopy (SEM) image in Figure 1d shows a smooth surface and stacked layers with clear edges, which illustrates prominent 2D-like features in m-GaTe. The evident cleavages and gaps between the layers hint at the weak vdW interaction in m-GaTe and a large feasibility of exfoliation into monolayers. We thus mechanically exfoliated a thin flake from the bulk sample transferred onto an $SiO_2$/Si substrate, with a sample thickness of ~20 nm, as shown by the atomic force microscopy (AFM) image in Figure 1e. Next, we chose the most prominent $A_g$ mode (at 115 cm$^{−1}$) and performed angle-resolved polarized Raman spectroscopy (ARPRS) on this ~20 nm thin film under 532 nm laser excitation. A substantial in-plane vibrational anisotropy can be identified, with a minimum and maximum intensity along two in-plane perpendicular directions (Figure 1f), consistent with previous work in ref. [17].

To more unambiguously identify the in-plane structural anisotropy, we carried out atomic-resolution STEM experiments on m-GaTe. In principle, the best direction for such characterization should be perpendicular to the $(2\bar{1}0)$ crystal plane, which however does not coincide with the zone axis in STEM. If the incident electron beam is not parallel to the zone axis, such deviation will result in over-complicated patterns and even a false image.[52-53] According to the Kikuchi bands, we tilted the sample to find the nearest zone axis that deviates from the normal of the $(2\bar{1}0)$ plane by roughly 4°, i.e. the $[1\bar{1}0]$ zone axis. The obtained high-angle annular dark field (HAADF) image is shown in **Figure 2**a. The intensity of the each HAADF spot is approximately proportional to $Z^2$, where $Z$ represents the averaged atomic number of the individual atomic column along the zone axis.[52] Hence, the brighter spots in the image indicate the Te columns whereas the weaker spots represent the Ga columns. We note that the typical building unit in the image is a hexagonal pattern (highlighted by red hexagons) with six bright corner points that represent the Te columns. Inside the hexagon, some faint spots are observed (Ga columns), which connect into a horizontal line-shaped pattern.

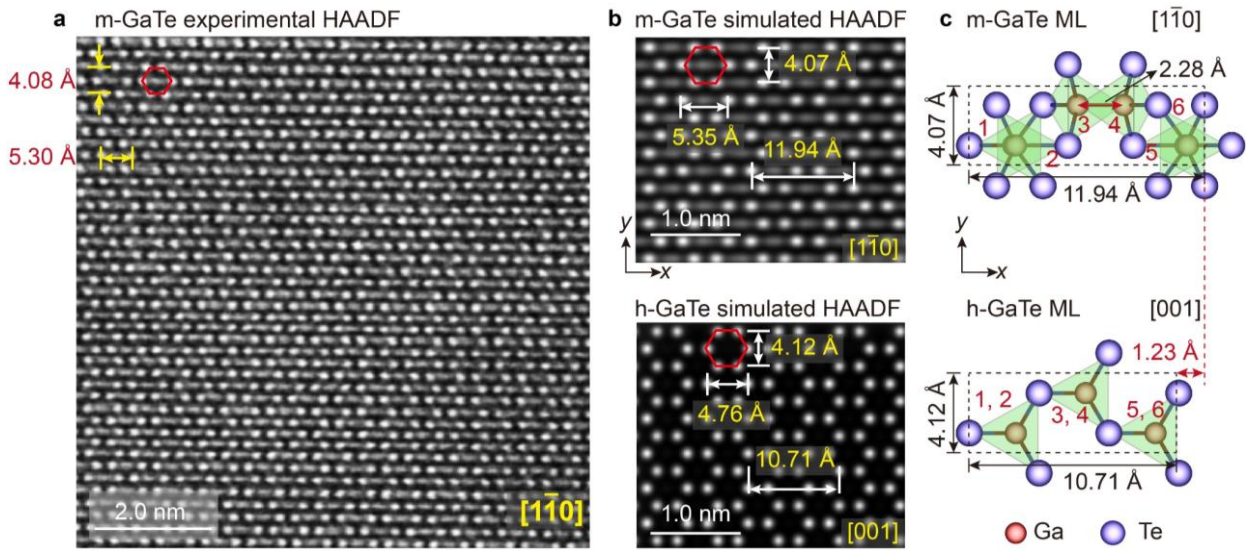

**Figure 2**. **Atomic-scale characterization of structural in-plane anisotropy**. a) The experimental atomic-resolution HAADF image of bulk m-GaTe viewed from the $[1\bar{1}0]$ zone axis. The six bright spots of the red hexagon are Te columns. b) The simulated HAADF images of bulk m-GaTe and h-GaTe. c) The atomic structure of the m-GaTe monolayer (ML) from the $[1\bar{1}0]$ zone axis and a 1×3 supercell of the h-GaTe monolayer from the top view. Ga[GaTe$_3$] tetrahedra are the building units of both structures, labelled as 1–6 in the images.

To enable a direct side-by-side comparison between the HAADF image and our DFT-relaxed structural model, we performed HAADF image simulations on the relaxed m-GaTe model with the view axis being set as [1$\bar{1}$0] as well as on the relaxed h-GaTe along the top view. The simulated HAADF image was obtained from a sample of ~20 nm thick, which showed great agreement with the experimental image (Figure 2b). The typical distances between atoms in the building unit are highly consistent between the experimentally measured and DFT-simulated HAADF images, viz. ~5.30 Å and ~4.08 Å (STEM experiments) versus 5.35 Å and 4.07 Å (simulated HAADF images). These atomic-scale images provide a direct real-space measure of the in-plane structural anisotropy in bulk m-GaTe along the $x$- and $y$-axis. To gain a clearer view of the in-plane structural anisotropy, we show one atomic slab of m-GaTe and h-GaTe (denoted as "ML", monolayer) in Figure 2c. In such a cell, both models contain six Ga-centered tetrahedra. The h-GaTe model shows an ordered arrangement of heteropolar bonds in the $x$-$y$ plane and thus has a considerable in-plane isotopy. By contrast, the appearance of tilted i-Ga–Ga homopolar bonds in m-GaTe breaks the rotational symmetry, resulting in a large expansion along the $x$-axis by 1.23 Å (and a marginal change in the $y$-axis by 0.05 Å), as well as a major structural difference, i.e., both Ga–Ga and Ga–Te bonds can be found along the $x$-axis but only Ga–Te bonds can be found along the $y$-axis. Hence, the formation of i-Ga–Ga homopolar bonds gives rise to the large structural in-plane anisotropy of m-GaTe.

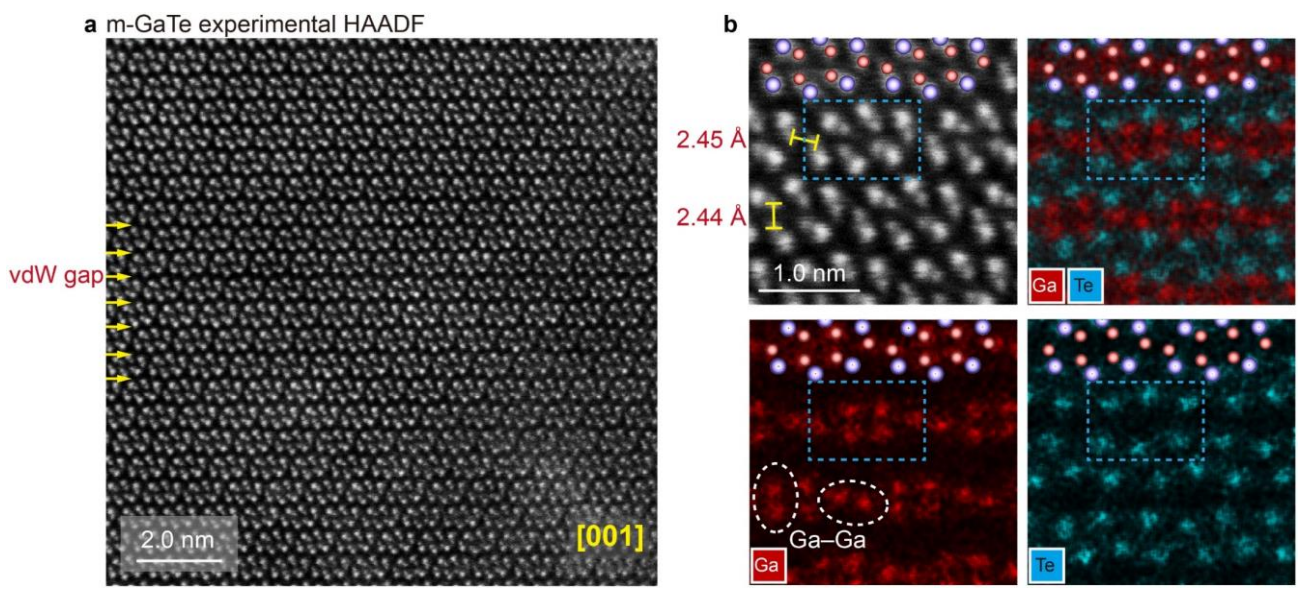

**Figure 3**. **Atomic-scale characterization of structural building units in anisotropic m-GaTe**. a) The atomic-resolution HAADF image of m-GaTe from the [001] zone axis. b) The HAADF image, with an overlaid map of Ga and Te atoms, respectively. The blue dashed-line boxes indicate the structural unit, and atoms forming Ga–Ga homopolar bonds are emphasized by white circles. The crystal structure images overlaid in the top region of each panel match very well with the experimental images.

To gain further insight of the homopolar bonds, we also recorded a HAADF image in the side view. According to the m-GaTe model shown in Figure 1a, each HAADF spot along the electron beam direction should correspond to an atomic column of only one element. As shown in **Figure 3**a, a prominent layered-like structure with visible zigzag-like vdW gaps is observed from the side view. To further clarify this point, we recorded the HAADF image in the [123] zone axis (Figure S2). The direction of vdW gaps is along the $(2\bar{1}0)$ plane, which is the same as that in Figure 3a. Within each slab, if we only look at the brighter Te columns, the structure seems to form pseudo-1D "nanowire" structures (cf. Figure 1a). However, these structural blocks are well connected by i-Ga–Ga bonds, with weaker image intensity of Ga-rich columns though. To better identify the important Ga–Ga homopolar bonds, we recorded the zoomed-in HAADF image and the corresponding EDX maps, shown in Figure 3b. The vertical Ga–Ga (v-Ga–Ga) bonds are also clearly identified in addition to the i-Ga–Ga bonds. These two types of Ga–Ga bonds show a nearly identical interatomic distance of ~2.45 Å. These experimental values agree very well with DFT data, i.e. ~2.45 Å, see **Figure 4**a. We overlaid images of the DFT-relaxed models at the top area of the zoomed-in HAADF images and EDX maps in Figure 4b, where the atomic positions are well matched.

To connect the structural anisotropy to practical applications, we calculated the optical properties of bulk m-GaTe and h-GaTe based on DFT calculations. The calculated real ($\varepsilon_1$) and imaginary ($\varepsilon_2$) parts of the dielectric function along the $x$-axis ($\varepsilon_{1,x}$, $\varepsilon_{2,x}$) and the $y$-axis ($\varepsilon_{1,y}$, $\varepsilon_{2,y}$) are shown in Figure S3a–b. We computed the refractive index ($n$) and extinction coefficient ($k$) using the following equations:[54]

$$n_i = \left( \frac{\sqrt{\varepsilon_{1,i}^2 + \varepsilon_{2,i}^2} + \varepsilon_{1,i}}{2} \right)^{\frac{1}{2}}, \quad (i=x, y) \qquad (1)$$

$$k_i = \left( \frac{\sqrt{\varepsilon_{1,i}^2 + \varepsilon_{2,i}^2} - \varepsilon_{1,i}}{2} \right)^{\frac{1}{2}}, \quad (i=x, y) \qquad (2)$$

A direct comparison of the polarization-dependent optical response $k$ between m- and h-GaTe is shown in Figure 4b and Figure 4c, and the result of the optical response $n$ is shown in Figure S3c. For m-GaTe, a considerable optical contrast was found between the $x$- and $y$-axis almost across the whole visible light region, i.e., from 300 to 800 nm of wavelength. By contrast, no anisotropy effects on optical profiles can be observed in h-GaTe. Although this anisotropic optical contrast is much smaller than what can be induced by true phase transitions, e.g. between amorphous and crystalline phases of GeTe and $Ge_2Sb_2Te_5$ phase-change materials (PCMs),[55-57] m-GaTe shows advantages in fast switching, high cycling endurance, and very limited power consumption, as this anisotropic optical switching only requires changing the polarization state of incident light.

To reveal the origin of such anisotropy in m-GaTe, we analyze the chemical bonding properties of the Ga–Ga bonds in both GaTe polymorphs using the crystal orbital Hamilton population (COHP)[58] method based on DFT-computed electronic wavefunctions. This method is useful to identify the orbital interactions and bonding nature of any interatomic contact in a wide range of material systems,[59] including complex crystalline solids[39] and amorphous materials.[60-62] As shown in Figure 4d, all Ga–Ga and Ga–Te contacts show strongly bonding interactions below the Fermi level $E_F$ (−COHP > 0, plotted to the right), indicating strong covalent bonds and good chemical stability of m-GaTe. As a link between two adjacent structural units in m-GaTe, the strong covalency of the i-Ga–Ga bonds underlines that m-GaTe should, in fact, be viewed as a 2D material: the building block is the complex 2D-like Te–Ga–Ga–Te quadruple-layers connected via i-Ga–Ga bonds, rather than a series of 1D nanowires. In fact, the experimental characterization of the stacked layer structure also supported the integrity of the 2D-like motifs in m-GaTe.[16] We carried out similar bonding analyses for h-GaTe, and both Ga–Te and v-Ga–Ga contacts in this polymorph were shown to be strong covalent bonds (Figure 4e).

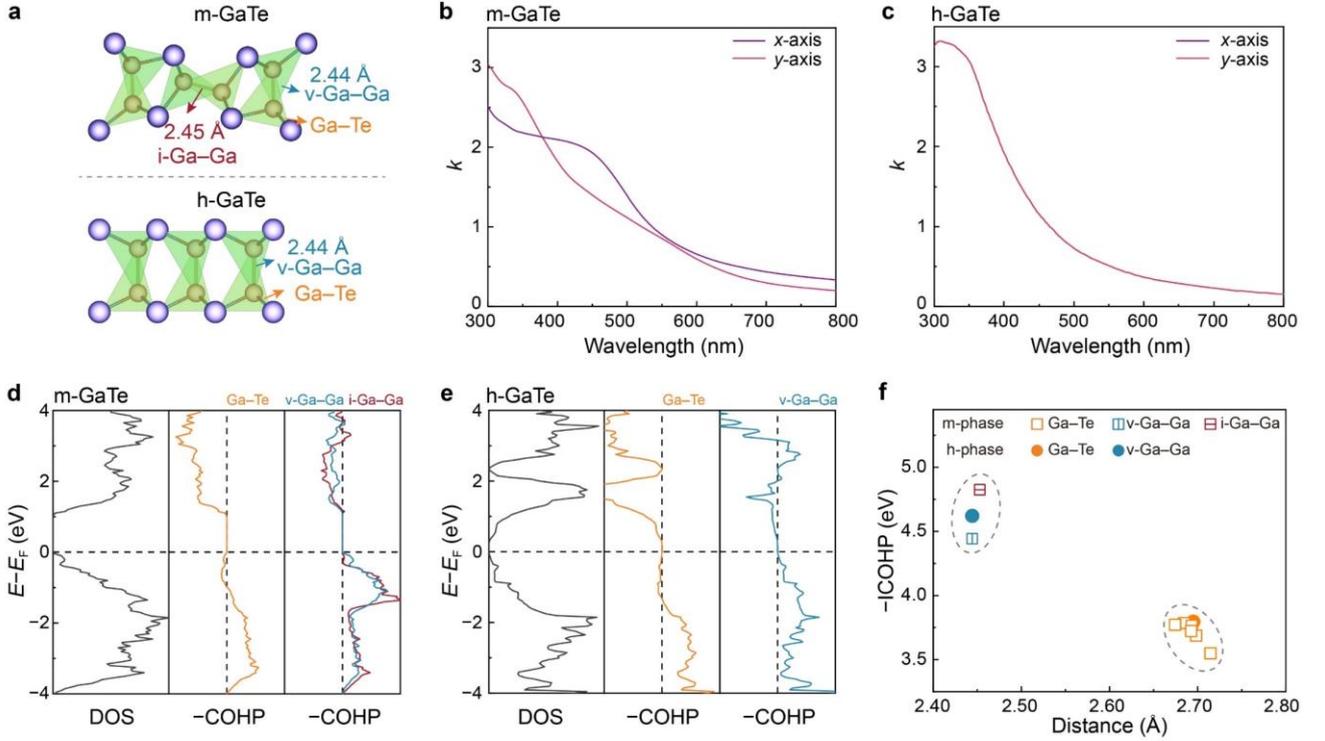

**Figure 4**. **Chemical-bonding nature and optical properties of GaTe from first principles**. a) Fragments from the crystal structures of m- and h- phase GaTe. The h-phase can be viewed as two intergrown Ga-centered tetrahedral motifs with opposite direction, sharing the same Ga–Ga bond, while one such tetrahedron is rotated by nearly 90° in the m-phase. b–c) The calculated along the *x*- and *y*-axis of bulk m-GaTe and h-GaTe, respectively. d–e) Computed electronic structures and chemical bonding interactions, as indicated by DOS and −COHP plots. f) Intralayer bonding analyses of h-GaTe and m-GaTe, showing the strength of individual interactions (measured by integrating the −COHP up to the Fermi level) plotted against the corresponding interatomic distances.

The integral of the −COHP data along the energy axis up to the Fermi level, $E_F$ (denoted as "−ICOHP" in the following) yields a quantitative measure of covalent interactions which may be correlated with the bond strength[47, 63] (Figure 4f). We find that all Ga–Ga bonds have shorter bond lengths and higher −ICOHP values compared to Ga–Te bonds for both m- and h-GaTe. The ICOHP-based bond strength of heteropolar Ga–Te bonds in m-GaTe is, in general, smaller than that of the Ga–Te bonds in h-GaTe. Albeit a direct comparison of integral values for different atomic species should be interpreted with care, the Ga–Ga homopolar bonds are crucial in stabilizing the Ga-centered tetrahedral motifs in GaTe. Interestingly, despite the similar bond length, the −ICOHP value of the i-Ga–Ga bond (~4.8 eV) is larger than that of the v-Ga–Ga bond in the m-phase (~4.4 eV) and in the h-phase (~4.6 eV). Given a possible transition from metastable h-GaTe to stable m-GaTe via rotating one of the vertical bonds by ~90°,[7] the formation of such distorted and strong i-Ga–Ga bonds gives rise to an increased chemical stability of the m-phase. A similar distortion-induced mechanism was previously found in monoclinic GeS$_2$,[39] in which only heteropolar Ge–S bonds are present. Upon tilting of the corner-shared GeS$_4$ tetrahedra along the *x*-axis in the compact segment of the monoclinic GeS$_2$ slabs, the chemical stability is increased as compared to the ordered tetragonal HgI$_2$-type phase.[39, 64] In contrast, the structural anisotropy in m-GaTe stems from tilted homopolar Ga–Ga bonds.

We note that the homopolar-induced anisotropy might be a genetic scheme in other vdW materials with similar local patterns as in m-GaTe. To test this idea, we performed a thorough materials screening over more than 154,000 known and hypothetical compounds in the Materials Project[65] via the following screening steps. We first set the chemical ratio as 1:1 and found more than 3,572 binary compounds. Next, we used the Rank Determination Algorithm (RDA) method[66] to single out 161 layer-structured binary materials. Then, we analyzed the local atomic environments using the *q* order parameter[67] and identified 35 structures that contain only tetrahedral motifs. Finally, based on a primitive ring analysis using the R.I.N.G.S code,[68] we picked up the structures that contain both 5- and 6-fold rings like m-GaTe. In total, five layered materials with distorted tetrahedral motifs were identified, viz. the monoclinic phases of GaTe, GeP, GeAs, SiAs, and the orthorhombic phase (o-) of SiP (**Figure 5**a). Figure 5b shows the monoclinic crystal structure, and the DFT-relaxed lattice parameters are included in Table S1. We note that these four IV–V low-symmetric crystals are experimentally available, and some of them have already enabled applications in high-performance photodetectors[69-72] and transistors.[73-75]

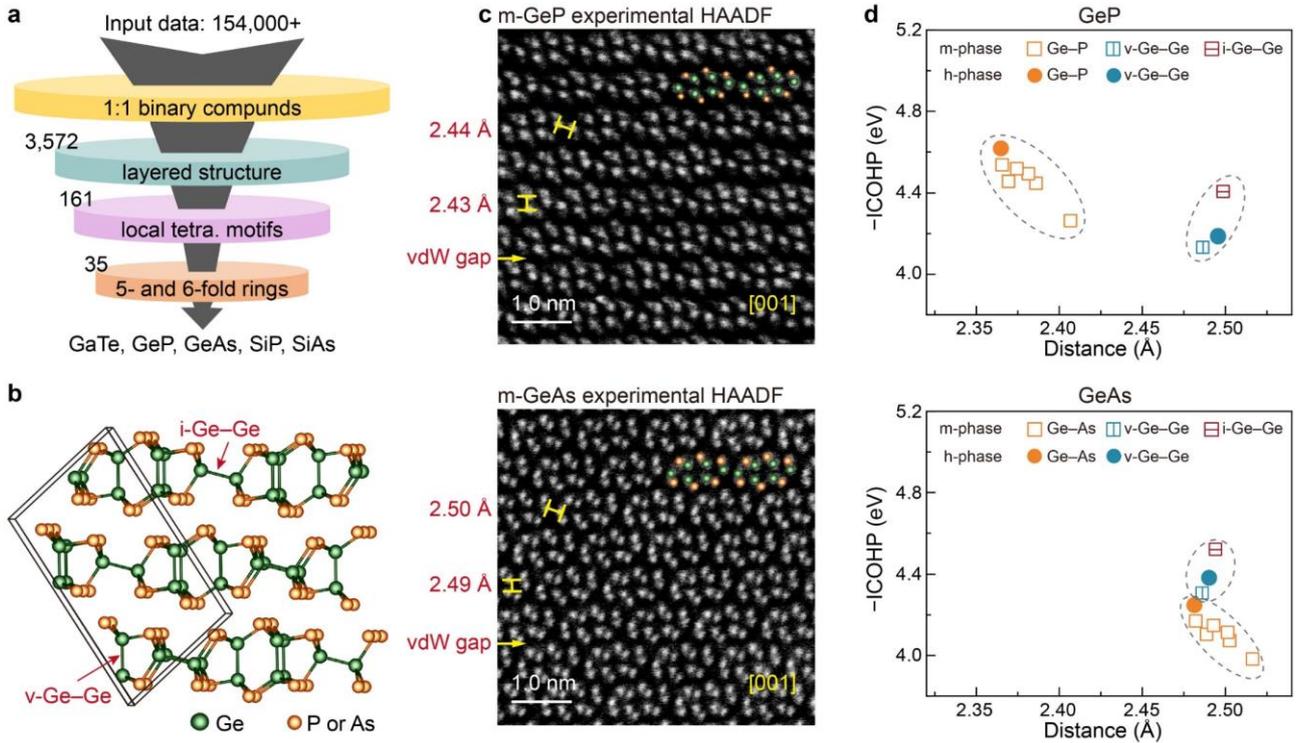

**Figure 5**. **Homopolar-bond-mediated anisotropy in related layered materials**. a) The flow of materials screening. b) The crystal structure of m-GeP and m-GeAs. c) The experimental HAADF images of m-GeP and m-GeAs viewed in the [001] zone axis. d) Chemical bonding analyses of GeP and GeAs.

We obtained two single-crystal samples of m-GeP and m-GeAs using the metallic Sn flux method and chemical vapor transport (CVT), respectively, and performed STEM-HAADF experiments on them. As shown in Figure 5c, their atomic arrangement is similar to m-GaTe in the [001] zone axis. In comparison with m-GaTe, the Ge–Ge homopolar bonds can be distinguished more clearly in the HAADF images, because the Ge columns show higher (or comparable) image intensity as compared to the P (or As) columns. We then quantified the degree of covalent interactions between the homopolar and heteropolar bonds in these disordered structures. We also considered the hypothetical hexagonal structures of GeP and GeAs with an ordered arrangement of tetrahedrons. As shown in Figure 5d, we found that all heteropolar bonds (Ge–P and Ge–As) in the low-symmetric structures are weaker (i.e., smaller −ICOHP) than the heteropolar bonds in their ordered phase. However, the −ICOHP value of i-Ge–Ge bonds in the low-symmetric phase is consistently larger than all v-Ge–Ge homopolar bonds in both m- and h-phase of GeP and GeAs. Similar bonding tendency is also found in m-SiAs and o-SiP (Figure S4). Note that the heteropolar bonds are systematically shorter with larger −ICOHP value than the homopolar bonds in m-GeP, which is clearly opposite to the case in m-GaTe. Hence, this enhanced chemical stability by the tilted in-plane homopolar bonds is valid regardless the length and strength of the corresponding heteropolar bonds, and explains why these layered materials tend to form disordered monoclinic and orthorhombic structures rather than an ordered hexagonal structure.

## 3. Conclusion

We have directly characterized the atomistic structure of the layered material m-GaTe with advanced microscopy techniques, providing imaging from two different view directions with sub-angstrom resolution and insight into the chemical identities of individual atoms. We observed the layered-like structures with clear vdW gaps and two different types of Ga–Ga homopolar bonds which are the crucial building blocks of Ga-centered tetrahedra. We quantified the large structural anisotropy of m-GaTe, mapping the complex crystal structure onto our DFT-relaxed models, and obtained a good agreement between our experimental and simulated results. We then carried out a comprehensive study of electronic structure and chemical bonding in m-GaTe. Strong covalent bonding was found not only for the heteropolar Ga–Te bonds, but also for the homopolar Ga–Ga bonds, which stabilize the structural building blocks in two-dimensional GaTe. We also found that the quantified bond strength of i-Ga–Ga bonds is larger than that of all Ga–Te bonds and v-Ga–Ga bonds in m-GaTe, emphasizing how important the i-Ga–Ga bonds are in determining the chemical stability of the material. Based on a high-throughput screening for local patterns similar to m-GaTe, we reveal a more general way of how tilted homopolar bonds mediate and stabilize the large in-plane anisotropy of this class of layered materials. Hence, based on experimentally verified atomic-resolution images and quantitative chemical bonding analysis, our work might guide the search for and experimental synthesis of more anisotropic layered materials. We also anticipate more advanced experimental tools, such as 4D-STEM[76] and correlative STEM and atom probe tomography (APT),[77] can be applied to gain further understanding of the chemical bonding characteristics of m-GaTe and related low-symmetric vdW crystals.

## 4. Methods

*Sample preparation and characterization*: Bulk GaTe and GeAs samples were grown using CVT, and GeP samples were grown by the metallic Sn flux method (these samples are commercially available from Six Carbon Technology Co., Ltd., Shenzhen and Nanjing MKNANO Tech. Co., Ltd). The monoclinic structure of the as-grown bulk GaTe samples was confirmed using XRD and Raman spectroscopy. The XRD experiments were carried out using a Bruker D8 Advance, Bruker AXS, with Cu K$\alpha$ source. The Raman spectroscopy experiments were performed in a Renishaw inVia Qontor at room temperature with an excitation laser wavelength of 532 nm. The SEM images were obtained with a Hitachi SU8230 SEM. The m-GaTe thin flakes were mechanically exfoliated from the bulk crystal using Scotch tape onto a Si substrate with 300 nm SiO$_2$. The thickness of the flakes was then identified using an AFM. In the polarization Raman measurements, the polarization direction of the incident laser light was rotated and the polar plots were fitted by the equation, $y = y_0 + A(\cos^2\theta)$. To characterize the structure in top view, bulk GaTe crystal materials were mechanically exfoliated into thin flakes, which were then transferred to a TEM grid (SiN grid). Focused ion beams (FIB, Hitachi NX5000) were used to fabricate cross-sectional TEM specimens, including GaTe, GeP, and GeAs. These TEM samples were characterized using Cs-corrected STEM (Hitachi HF5000) at 200 kV with a 0.052 nA beam current, and we then obtained the HAADF and EDX mapping images in a detector range of 40-213 mrad. The total counts of EDX mapping images were ~50,000. The element-resolved atomic-scale imaging approach plays a key role in understanding complex crystalline structures and defects.[52, 78-81] The HAADF image simulations were performed using Dr. Probe packages,[82] in which the acceleration voltage was set to be 200 kV, the aperture radius alpha was set to be 25 mrad, and the detection range of the HAADF detector was set to be 40-213 mrad. The sample thickness for the HAADF image simulation was set to be 20 nm.

*Ab initio calculations*: DFT calculations were carried out using the Vienna Ab Initio Simulation Package (VASP)[45] with projector augmented wave (PAW) pseudopotentials.[41-42] Crystal structures of both m- and h-GaTe were fully relaxed with respect to both atomic coordinates and cell volumes. The Perdew–Burke–Ernzerhof (PBE) functional[43] and the Grimme D3 method[44] for vdW interactions were employed for structural optimizations, electronic structure calculations and chemical bonding analyses. The chemical bonding analyses based on crystal orbital Hamilton populations (COHP)[58] were carried out using the Local Orbital Basis Suite Towards Electronic-Structure Reconstruction (LOBSTER) code,[83-85] which projects the self-consistent wavefunction into an auxiliary basis of local orbitals, thereby giving access to projected COHP analysis. The cut-off energy for plane waves was set to 500 eV. A 3 × 5 × 12 *k*-point mesh was used for bulk monoclinic and orthorhombic models, and 12 × 12 × 3 *k*-point meshes were used for bulk hexagonal models. Atomic structures were visualized using VESTA.[86]

**Conflict of Interest**
The authors declare no competing interests.

**Supporting Information**
Supporting Information is available from …

**Acknowledgements**

The authors thank Chaobin Zeng from Hitachi High-Tech Scientific Solutions (Beijing) Co., Ltd. and Yuanbin Qin for technical support on STEM experiments. We acknowledge the Shenzhen 6Carbon Technology Co., Ltd. and Nanjing MKNANO Tech. Co., Ltd. for providing the single crystal samples. We thank Dr. Xudong Wang for his preliminary efforts on DFT calculations. The authors acknowledge the support of the International Joint Laboratory for Micro/Nano Manufacturing and Measurement Technologies and the HPC platform of Xi'an Jiaotong University and the Computing Center in Xi'an.

**Data Availability Statement**

Data supporting this work will be available at https://caid.xjtu.edu.cn/info upon journal publication.

**References**

[1] S. Siddique, C. C. Gowda, R. Tromer, S. Demiss, A. R. S. Gautam, O. E. Femi, P. Kumbhakar, D. S. Galvao, A. Chandra, C. S. Tiwary, *ACS Appl. Nano Mater.* **2021**, 4, 4829.

[2] F. Bondino, S. Duman, S. Nappini, G. D'Olimpio, C. Ghica, T. O. Menteş, F. Mazzola, M. C. Istrate, M. Jugovac, M. Vorokhta, S. Santoro, B. Gürbulak, A. Locatelli, D. W. Boukhvalov, A. Politano, *Adv. Funct. Mater.* **2022**, 32, 2205923.

[3] F. Liu, H. Shimotani, H. Shang, T. Kanagasekaran, V. Zolyomi, N. Drummond, V. I. Fal'ko, K. Tanigaki, *ACS Nano* **2014**, 8, 752.

[4] P. Hu, J. Zhang, M. Yoon, X.-F. Qiao, X. Zhang, W. Feng, P. Tan, W. Zheng, J. Liu, X. Wang, J. C. Idrobo, D. B. Geohegan, K. Xiao, *Nano Res.* **2014**, 7, 694.

[5] M. Julien-Pouzol, S. Jaulmes, M. Guittard, F. Alapini, *Acta Crystallogr. Sect. B* **1979**, 35, 2848.

[6] S. Semiletov, V. Vlasov, *Kristallografiya* **1963**, 8, 877.

[7] Q. Zhao, T. Wang, Y. Miao, F. Ma, Y. Xie, X. Ma, Y. Gu, J. Li, J. He, B. Chen, S. Xi, L. Xu, H. Zhen, Z. Yin, J. Li, J. Ren, W. Jie, *Phys. Chem. Chem. Phys.* **2016**, 18, 18719.

[8] E. Zallo, A. Pianetti, A. S. Prikhodko, S. Cecchi, Y. S. Zaytseva, A. Giuliani, M. Kremser, N. I. Borgardt, J. J. Finley, F. Arciprete, M. Palummo, O. Pulci, R. Calarco, *npj 2D Mater. Appl.* **2023**, 7, 19.

[9] Y. Yu, M. Ran, S. Zhou, R. Wang, F. Zhou, H. Li, L. Gan, M. Zhu, T. Zhai, *Adv. Funct. Mater.* **2019**, 29, 1901012.

[10] M. Liu, S. Yang, M. Han, S. Feng, G. G. Wang, L. Dang, B. Zou, Y. Cai, H. Sun, J. Yu, J. C. Han, Z. Liu, *Small* **2021**, 17, 2007909.

[11] X. Han, *Science* **2020**, 369, 509.

[12] X.-D. Wang, J. Tan, J. Ouyang, H.-M. Zhang, J.-J. Wang, Y. Wang, V. L. Deringer, J. Zhou, W. Zhang, E. Ma, *Adv. Sci.* **2022**, 9, 2203776.

[13] Z. Gao, T.-R. Wei, T. Deng, P. Qiu, W. Xu, Y. Wang, L. Chen, X. Shi, *Nat. Commun.* **2022**, 13, 7491.

[14] T.-R. Wei, M. Jin, Y. Wang, H. Chen, Z. Gao, K. Zhao, P. Qiu, Z. Shan, J. Jiang, R. Li, L. Chen, J. He, X. Shi, *Science* **2020**, 369, 542.

[15] H. Huang, H. Chen, Z. Gao, Y. Ma, K. Zhao, T.-R. Wei, X. Shi, *Adv. Funct. Mater.* **2023**, 33, 2306042.

[16] H. Cai, B. Chen, G. Wang, E. Soignard, A. Khosravi, M. Manca, X. Marie, S. L. Chang, B. Urbaszek, S. Tongay, *Adv. Mater.* **2017**, 29, 1605551.

[17] S. Huang, Y. Tatsumi, X. Ling, H. Guo, Z. Wang, G. Watson, A. A. Puretzky, D. B. Geohegan, J. Kong, J. Li, T. Yang, R. Saito, M. S. Dresselhaus, *ACS Nano* **2016**, 10, 8964.

[18] M. Tong, Y. Hu, W. He, X.-L. Yu, S. Hu, X. a. Cheng, T. Jiang, *ACS Nano* **2021**, 15, 17565.

[19] H. Wang, M. L. Chen, M. Zhu, Y. Wang, B. Dong, X. Sun, X. Zhang, S. Cao, X. Li, J. Huang, L. Zhang, W. Liu, D. Sun, Y. Ye, K. Song, J. Wang, Y. Han, T. Yang, H. Guo, C. Qin, L. Xiao, J. Zhang, J. Chen, Z. Han, Z. Zhang, *Nat. Commun.* **2019**, 10, 2302.

[20] A. Yamamoto, A. Syouji, T. Goto, E. Kulatov, K. Ohno, Y. Kawazoe, K. Uchida, N. Miura, *Phys. Rev. B* **2001**, 64, 035210.

[21] Z. Wang, M. Safdar, M. Mirza, K. Xu, Q. Wang, Y. Huang, F. Wang, X. Zhan, J. He, *Nanoscale* **2015**, 7, 7252.

[22] J. Zhou, T. Xiong, Z. Guo, K. Xin, X. Wang, H. Gu, Y. Y. Liu, L. Liu, J. Yang, Z. Wei, *IEEE Trans. Electron Devices* **2023**, 70, 1715.

[23] Z. Wang, K. Xu, Y. Li, X. Zhan, M. Safdar, Q. Wang, F. Wang, J. He, *ACS Nano* **2014**, 8, 4859.

[24] C. C. Sta. Maria, P.-H. Wu, D. P. Hasibuan, C. S. Saragih, H. Giap, D. H. Nguyen, Y.-R. Chen, R. A. Patil, D. V. Pham, J.-L. Shen, C.-C. Lai, M.-K. Wu, Y.-R. Ma, *J. Mater. Chem. C* **2023**, 11, 14316.

[25] J. Kang, V. K. Sangwan, H.-S. Lee, X. Liu, M. C. Hersam, *ACS Photonics* **2018**, 5, 3996.

[26] H. Liu, A. T. Neal, Z. Zhu, Z. Luo, X. Xu, D. Tománek, P. D. Ye, *ACS Nano* **2014**, 8, 4033.

[27] L. Li, Y. Yu, G. J. Ye, Q. Ge, X. Ou, H. Wu, D. Feng, X. H. Chen, Y. Zhang, *Nat. Nanotechnol.* **2014**, 9, 372.

[28] X. Li, H. Liu, C. Ke, W. Tang, M. Liu, F. Huang, Y. Wu, Z. Wu, J. Kang, *Laser Photonics Rev.* **2021**, 15, 2100322.

[29] M. Tang, B. Wang, H. Lou, F. Li, A. Bergara, G. Yang, *J. Phys. Chem. Lett.* **2021**, 12, 8320.

[30] W. Yang, K. Xin, J. Yang, Q. Xu, C. Shan, Z. Wei, *Small Methods* **2022**, 6, 2101348.

[31] Y. Yang, S.-C. Liu, Z. Li, D.-J. Xue, J.-S. Hu, *Chem. Commun.* **2021**, 57, 565.

[32] Y. Wang, Y. Zhao, X. Ding, L. Qiao, *J. Energy Chem.* **2021**, 60, 451.

[33] J. Susoma, L. Karvonen, A. Säynätjoki, S. Mehravar, R. A. Norwood, N. Peyghambarian, K. Kieu, H. Lipsanen, J. Riikonen, *Appl. Phys. Lett.* **2016**, 108, 073103.

[34] L. Li, W. Han, L. Pi, P. Niu, J. Han, C. Wang, B. Su, H. Li, J. Xiong, Y. Bando, T. Zhai, *InfoMat* **2019**, 1, 54.

[35] F. Zhao, Y. Feng, W. Feng, *InfoMat* **2022**, 4, e12365.

[36] S. Zhao, P. Luo, S. Yang, X. Zhou, Z. Wang, C. Li, S. Wang, T. Zhai, X. Tao, *Adv. Opt. Mater.* **2021**, 9, 2100198.

[37] C. Zhang, H. Ouyang, R. Miao, Y. Sui, H. Hao, Y. Tang, J. You, X. Zheng, Z. Xu, X. a. Cheng, T. Jiang, *Adv. Opt. Mater.* **2019**, 7, 1900631.

[38] J. Zhao, D. Ma, C. Wang, Z. Guo, B. Zhang, J. Li, G. Nie, N. Xie, H. Zhang, *Nano Res.* **2021**, 14, 897.

[39] X. Wang, J. Tan, C. Han, J. J. Wang, L. Lu, H. Du, C. L. Jia, V. L. Deringer, J. Zhou, W. Zhang, *ACS Nano* **2020**, 14, 4456.

[40] Y. Yang, S. C. Liu, X. Wang, Z. Li, Y. Zhang, G. Zhang, D. J. Xue, J. S. Hu, *Adv. Funct. Mater.* **2019**, 29, 1900411.

[41] P. E. Blöchl, *Phys. Rev. B* **1994**, 50, 17953.

[42]   G. Kresse, D. Joubert, *Phys. Rev. B* **1999**, 59, 1758.

[43]   J. P. Perdew, K. Burke, M. Ernzerhof, *Phys. Rev. Lett.* **1996**, 77, 3865.

[44]   S. Grimme, J. Antony, S. Ehrlich, H. Krieg, *J. Chem. Phys.* **2010**, 132, 154104

[45]   G. Kresse, J. Hafner, *Phys. Rev. B* **1993**, 47, 558.

[46]   E. G. Gillan, A. R. Barron, *Chem. Mater.* **1997**, 9, 3037.

[47]   V. L. Deringer, R. P. Stoffel, M. Wuttig, R. Dronskowski, *Chem. Sci.* **2015**, 6, 5255.

[48]   Y. Du, G. Qiu, Y. Wang, M. Si, X. Xu, W. Wu, P. D. Ye, *Nano Lett.* **2017**, 17, 3965.

[49]   W. Wu, G. Qiu, Y. Wang, R. Wang, P. Ye, *Chem. Soc. Rev.* **2018**, 47, 7203.

[50]   C. J. Bae, J. McMahon, H. Detz, G. Strasser, J. Park, E. Einarsson, D. B. Eason, *AIP Adv.* **2017**, 7, 035113.

[51]   Z. Li, B. Xu, D. Liang, A. Pan, *Research* **2020**, 2020, 5464258.

[52]   S. J. Pennycook, P. D. Nellist, *Scanning Transmission Electron Microscopy: Imaging and Analysis*, Springer Science & Business Media, **2011**.

[53]   H. Sha, J. Cui, R. Yu, *Sci. Adv.* **2022**, 8, eabn2275.

[54]   V. Wang, N. Xu, J.-C. Liu, G. Tang, W.-T. Geng, *Comput. Phys. Commun.* **2021**, 267, 108033.

[55]   W. Zhang, R. Mazzarello, M. Wuttig, E. Ma, *Nat. Rev. Mater.* **2019**, 4, 150.

[56]   D. Wang, L. Zhao, S. Yu, X. Shen, J.-J. Wang, C. Hu, W. Zhou, W. Zhang, *Mater. Today* **2023**, 68, 334.

[57]   W. Zhou, X. Shen, X. Yang, J. Wang, W. Zhang, *Int. J. Extrem. Manuf.* **2024**, 6, 022001.

[58]   R. Dronskowski, P. E. Blöchl, *J. Phys. Chem.* **1993**, 97, 8617.

[59]   A. A. Naik, C. Ertural, N. Dhamrait, P. Benner, J. George, *Sci. Data* **2023**, 10, 610.

[60]   V. L. Deringer, W. Zhang, M. Lumeij, S. Maintz, M. Wuttig, R. Mazzarello, R. Dronskowski, *Angew. Chem. Int. Ed.* **2014**, 53, 10817.

[61]   F. Rao, K. Ding, Y. Zhou, Y. Zheng, M. Xia, S. Lv, Z. Song, S. Feng, I. Ronneberger, R. Mazzarello, W. Zhang, E. Ma, *Science* **2017**, 358, 1423.

[62]   L. Sun, Y. Zhou, X. Wang, Y. Chen, V. L. Deringer, R. Mazzarello, W. Zhang, *npj Comput. Mater.* **2021**, 7, 29.

[63]   M. Küpers, P. M. Konze, S. Maintz, S. Steinberg, A. M. Mio, O. Cojocaru-Miredin, M. Zhu, M. Müller, M. Luysberg, J. Mayer, M. Wuttig, R. Dronskowski, *Angew. Chem. Int. Ed.* **2017**, 56, 10204.

[64]   M. Shimada, F. Dachille, *Inorg. Chem.* **1977**, 16, 2094.

[65]   A. Jain, S. P. Ong, G. Hautier, W. Chen, W. D. Richards, S. Dacek, S. Cholia, D. Gunter, D. Skinner, G. Ceder, K. A. Persson, *APL Mater.* **2013**, 1, 011002.

[66]   P. M. Larsen, M. Pandey, M. Strange, K. W. Jacobsen, *Phys. Rev. Mater.* **2019**, 3, 034003.

[67]   P. Zalden, F. Quirin, M. Schumacher, J. Siegel, S. Wei, A. Koc, M. Nicoul, M. Trigo, P. Andreasson, H. Enquist, M. J. Shu, T. Pardini, M. Chollet, D. Zhu, H. Lemke, I. Ronneberger, J. Larsson, A. M. Lindenberg, H. E. Fischer, S. Hau-Riege, D. A. Reis, R. Mazzarello, M. Wuttig, K. Sokolowski-Tinten, *Science* **2019**, 364, 1062.

[68]   S. Le Roux, P. Jund, *Computational Materials Science* **2010**, 49, 70.

[69]   L. Li, W. Wang, P. Gong, X. Zhu, B. Deng, X. Shi, G. Gao, H. Li, T. Zhai, *Adv. Mater.* **2018**, 30, 1706771.

[70]   S. Yang, Y. Yang, M. Wu, C. Hu, W. Shen, Y. Gong, L. Huang, C. Jiang, Y. Zhang, P. M. Ajayan,

*Adv. Funct. Mater.* **2018**, 28, 1707379.

[71] S. Hou, Z. Guo, T. Xiong, X. Wang, J. Yang, Y.-Y. Liu, Z.-C. Niu, S. Liu, B. Liu, S. Zhai, H. Gu, Z. Wei, *Nano Res.* **2022**, 15, 8579.

[72] D. Kim, K. Park, J. H. Lee, I. S. Kwon, I. H. Kwak, J. Park, *Small* **2021**, 17, 2006310.

[73] F. Zou, Y. Cong, W. Song, H. Liu, Y. Li, Y. Zhu, Y. Zhao, Y. Pan, Q. Li, *Nanomaterials* **2024**, 14, 238.

[74] A. Grillo, A. Di Bartolomeo, F. Urban, M. Passacantando, J. M. Caridad, J. Sun, L. Camilli, *ACS Appl. Mater. Interfaces* **2020**, 12, 12998.

[75] J. Guo, Y. Liu, Y. Ma, E. Zhu, S. Lee, Z. Lu, Z. Zhao, C. Xu, S.-J. Lee, H. Wu, K. Kovnir, Y. Huang, X. Duan, *Adv. Mater.* **2018**, 30, 1705934.

[76] W. Gao, C. Addiego, H. Wang, X. Yan, Y. Hou, D. Ji, C. Heikes, Y. Zhang, L. Li, H. Huyan, T. Blum, T. Aoki, Y. Nie, D. G. Schlom, R. Wu, X. Pan, *Nature* **2019**, 575, 480.

[77] Y. Yu, C. Zhou, S. Zhang, M. Zhu, M. Wuttig, C. Scheu, D. Raabe, G. J. Snyder, B. Gault, O. Cojocaru-Mirédin, *Mater. Today* **2020**, 32, 260.

[78] B. Zhang, W. Zhang, Z. Shen, Y. Chen, J. Li, S. Zhang, Z. Zhang, M. Wuttig, R. Mazzarello, E. Ma, X. Han, *Appl. Phys. Lett.* **2016**, 108, 191902.

[79] J.-J. Wang, J. Wang, H. Du, L. Lu, P. C. Schmitz, J. Reindl, A. M. Mio, C.-L. Jia, E. Ma, R. Mazzarello, M. Wuttig, W. Zhang, *Chem. Mater.* **2018**, 30, 4770.

[80] J. Xu, J. He, Y. Ding, J. Luo, *Sci. China Mater.* **2020**, 63, 1788.

[81] J. Xu, X.-X. Xue, G. Shao, C. Jing, S. Dai, K. He, P. Jia, S. Wang, Y. Yuan, J. Luo, J. Lu, *Nat. Commun.* **2023**, 14, 7849.

[82] J. Barthel, *Ultramicroscopy* **2018**, 193, 1.

[83] V. L. Deringer, A. L. Tchougreeff, R. Dronskowski, *J. Phys. Chem. A.* **2011**, 115, 5461.

[84] S. Maintz, V. L. Deringer, A. L. Tchougréeff, R. Dronskowski, *J. Comput. Chem.* **2013**, 34, 2557.

[85] R. Nelson, C. Ertural, J. George, V. L. Deringer, G. Hautier, R. Dronskowski, *J. Comput. Chem.* **2020**, 41, 1931.

[86] K. Momma, F. Izumi, *J. Appl. Crystallogr.* **2008**, 41, 653.

We perform a comprehensive atomic-scale characterization on monoclinic gallium telluride (m-GaTe) via element-resolved atomic imaging, and reveal how Ga–Ga homopolar bonds stabilize the in-plane structural anisotropy of m-GaTe based on the clarified elemental information and quantum chemistry bonding.

*Jieling Tan[1], Jiang-Jing Wang[1]\*, Hang-Ming Zhang[1], Han-Yi Zhang[1], Heming Li[1,2], Yu Wang[2], Yuxing Zhou[3], Volker L. Deringer[3]\*, Wei Zhang[1]\**

# Homopolar Chemical Bonds Induce In-Plane Anisotropy in Layered Semiconductors

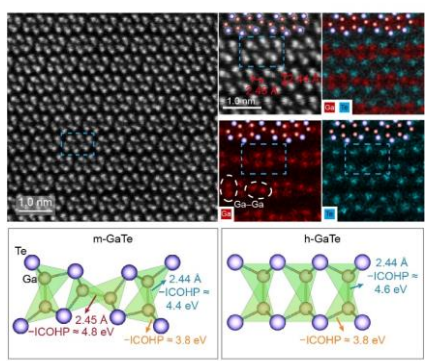

# Supporting Information

# Homopolar Chemical Bonds Induce In-Plane Anisotropy

# in Layered Semiconductors

*Jieling Tan[1], Jiang-Jing Wang[1]\*, Hang-Ming Zhang[1], Han-Yi Zhang[1], Heming Li[1,2], Yu Wang[2], Yuxing Zhou[3], Volker L. Deringer[3]\*, Wei Zhang[1]\**

[1]Center for Alloy Innovation and Design (CAID), State Key Laboratory for Mechanical Behavior of Materials, Xi'an Jiaotong University, Xi'an, 710049, China

[2]School of Physics, Xi'an Jiaotong University, Xi'an, 710049, China

[3]Inorganic Chemistry Laboratory, Department of Chemistry, University of Oxford, Oxford, OX1 3QR, UK

\*Emails: j.wang@mail.xjtu.edu.cn, volker.deringer@chem.ox.ac.uk, wzhang0@mail.xjtu.edu.cn

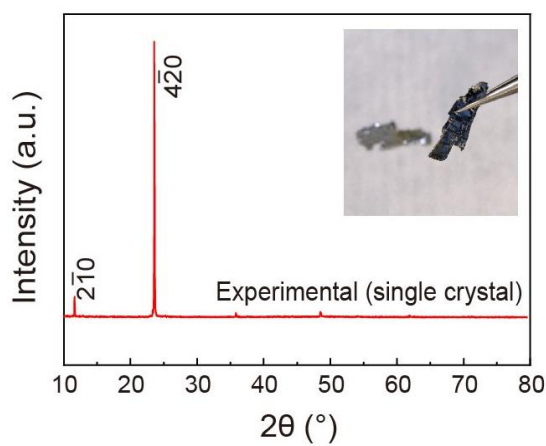

**Figure S1**. Experimental XRD pattern of bulk m-GaTe. The inset shows an optical image of the bulk sample. There are two strong peaks corresponding to the $(2\bar{1}0)$ and $(4\bar{2}0)$ planes.

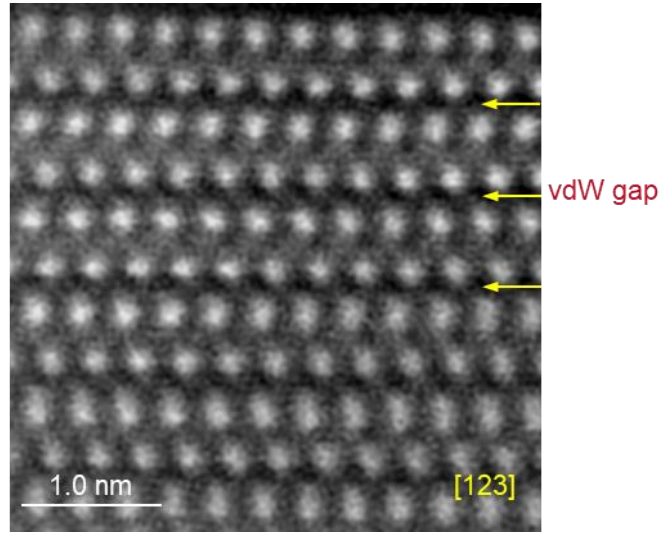

**Figure S2**. The recorded HAADF image in the [123] zone axis of m-GaTe.

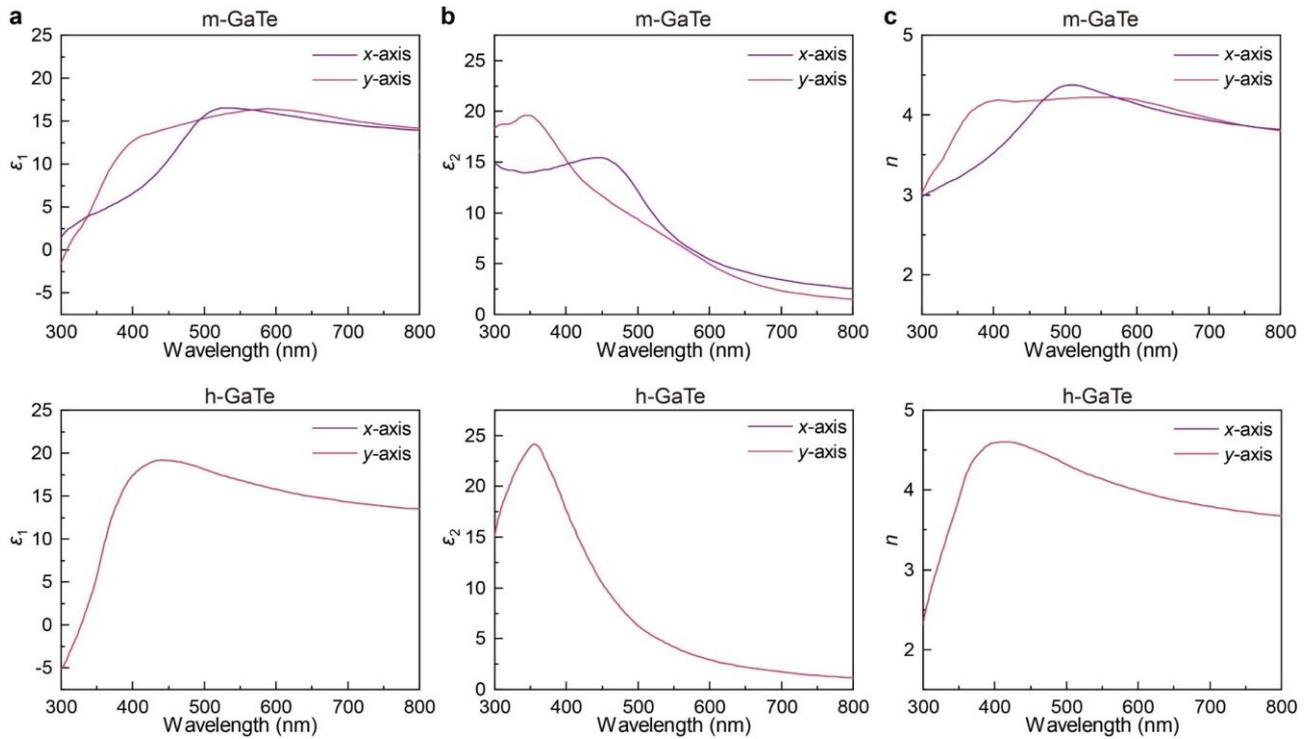

**Figure S3**. The DFT calculated $\varepsilon_1$, $\varepsilon_2$ and $n$ profiles of bulk m-GaTe (top) and h-GaTe (bottom) along the *x*- and *y*-axis, respectively.

**Table S1**. DFT-computed lattice parameters of the four IV–V compounds in monoclinic ("m-"), hexagonal ("h-"), and orthorhombic ("o-") phases. The four disordered crystals are experimentally available, but the ordered hexagonal structures are hypothetical.

|         | Lattice parameters | | | | |
|---------|--------|--------|--------|------------------------|--------|
|         | $a$ [Å] | $b$ [Å] | $c$ [Å] | $\alpha = \beta$ [°] | $\gamma$ [°] |
| m-GeP   | 15.17  | 9.21   | 3.68   | 90                     | 100.37 |
| h-GeP   | 3.65   | 3.65   | 15.63  | 90                     | 120    |
| m-GeAs  | 15.60  | 9.56   | 3.86   | 90                     | 100.33 |
| h-GeAs  | 3.82   | 3.82   | 15.76  | 90                     | 120    |
| o-SiP   | 20.51  | 13.79  | 3.52   | 90                     | 90     |
| h-SiP   | 3.52   | 3.52   | 15.42  | 90                     | 120    |
| m-SiAs  | 16.24  | 9.62   | 3.69   | 90                     | 106.60 |
| h-SiAs  | 3.68   | 3.68   | 15.82  | 90                     | 120    |

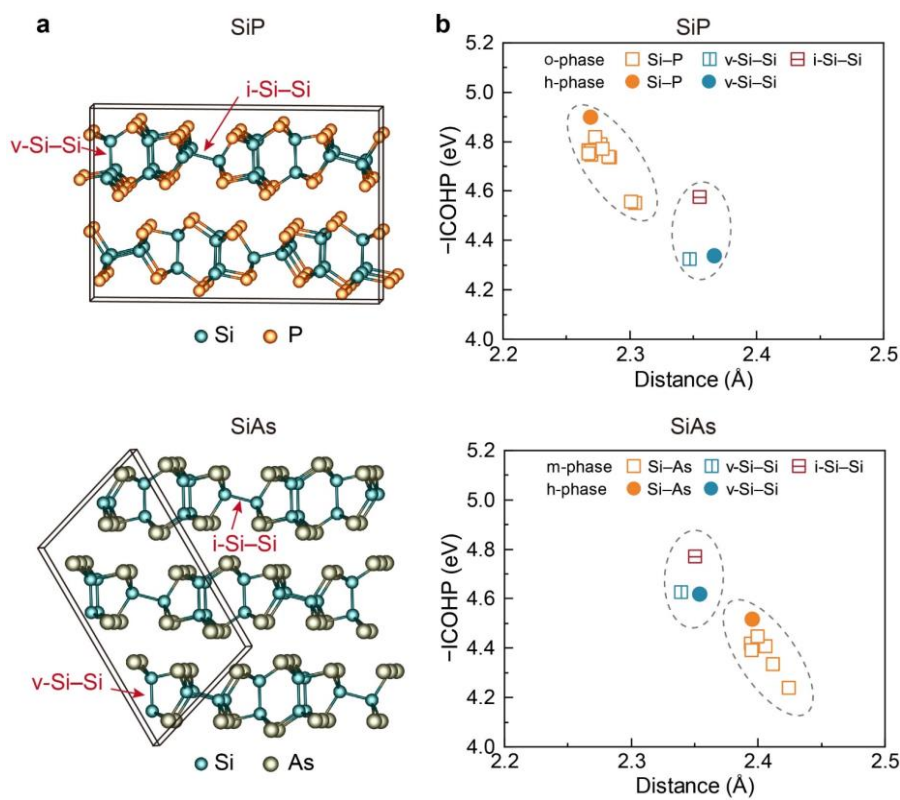

**Figure S4**. a) The crystal structures of o-SiP (top) and m-SiAs (bottom). b) The chemical bonding analyses of SiP and SiAs based on −ICOHP.